\documentclass[journal=jacsat,manuscript=article]{achemso}

\usepackage[version=3]{mhchem} 
\usepackage[utf8]{inputenc}
\usepackage[T1]{fontenc}
\usepackage{multirow}
\usepackage{caption}
\usepackage{listings}
\usepackage{amsmath}

\author{Daniel Maldonado-Lopez  }
\author{Noboru Takeuchi}
\author{Jonathan Guerrero-Sanchez}
\email{guerrero@cnyn.unam.mx}
\affiliation[Universidad Nacional Autónoma de México]
{Centro de Nanociencia y Nanotecnología, Universidad Nacional Autónoma de México, Ensenada, Baja California, Código Postal 22800, México}

\title[IrMn3Fe]
  {Understanding the non-collinear antiferromagnetic IrMn\textsubscript{3} surfaces and their exchange-biased heterostructures from first principles}

\keywords{Exchange bias, non-collinear magnetism, interfaces, surfaces, magnetic coupling}

\begin{document}



\newpage

\begin{abstract}
We provide a complete and systematic first-principles study of the thermodynamic stability, structural parameters, and magnetic properties of the T1 non-collinear antiferromagnetic L1\textsubscript{2}-IrMn\textsubscript{3} surface and L1\textsubscript{2}-IrMn\textsubscript{3}/Fe heterostructure. Furthermore, we investigate the exchange-bias effect in the heterostructure and describe its previously unknown complex magnetic coupling at the interface. We consider four atomic configurations and four magnetic arrangements, finding two stable terminations for the surface and for the heterostructure, which are in good agreement with experimental HR-TEM data. Using a comparative approach to analyze the exchange-bias properties of the heterostructure, we discover that the number of Mn-Fe interactions is related to the exchange bias intensity. This finding could lead to novel exchange-bias tailoring methods by controlling the termination of the layers. Finally, we are able to accurately describe the interface magnetic coupling atom-by-atom, and we find a relationship between the antiferromagnetic order at the interface and the stability of the heterostructure. Our analysis provides a possible mechanism for the appearance of exchange bias in non-collinear/collinear heterostructures, and it is in good agreement with experimental hysteresis measurements of the IrMn\textsubscript{3}/Fe system.
\end{abstract}

\section{Introduction}
\noindent The exchange bias (EB) effect, present in magnetically coupled antiferromagnetic/ferromagnetic (AF/FM) interfaces\cite{nogues1999exchange}, was discovered by W. H. Meiklejohn and C. P. Bean in 1956\cite{meiklejohn1956new}. They found a horizontal shift in the hysteresis loop of the Co-CoO system. This characteristic hysteresis loop of EB systems is measured with an exchange field ($H_E$), and it is usually accompanied by high coercitivities\cite{stiles2001coercivity} ($H_C$). The EB effect is applied in modern technologies, typically with the goal of pinning the magnetization of ferromagnetic layers near the AF/FM interface\cite{fernandez2008large}. Such technologies include microactuators\cite{pan2005magnetically}, and giant magnetoresistive devices\cite{coehoorn2000giant}, such as read heads in hard drives and Magnetic Random Access Memories, where the AF/FM interaction is used to control the inversion of ferromagnetic layers near the interface.

Exchange-biased systems have recently been of great interest due to promising applications in spintronics, including: sensors\cite{doi:10.1063/1.1450050, doi:10.1063/1.112304}, tunnel magnetoresistance devices\cite{doi:10.1063/1.2402204}, spin valves\cite{doi:10.1063/1.372907}, magnetic tunnel junctions\cite{fernandez2008large}, and spintronic memories\cite{4160113}. In order to better predict, characterize, and exploit the properties of heterostructures with EB, it is essential to describe the interface formation and interface magnetic coupling. Although there have been great advancements in the experimental techniques used to measure magnetic properties, e.g. neutron scattering, it is not yet feasible to measure the magnetic moment of individual atoms at the interface. Thus, it is difficult to accurately describe, and understand the processes that give rise to the EB effect. The problem of describing magnetic moments at the interface can be addressed through computational methods such as Density Functional Theory (DFT), which allows us to predict the magnetic behavior of materials atom-by-atom.

The choice of AF material used in exchange-biased heterostructures is crucial to increase the stability and EB properties at the interface. A promising AF material for exchange bias applications is IrMn\textsubscript{3}. This material can be found in two crystallographic phases: L1\textsubscript{2}-IrMn\textsubscript{3} and $\gamma$-IrMn\textsubscript{3}. In the present work, we focus on the L1\textsubscript{2} phase of IrMn\textsubscript{3}, since it has been proven that the exchange interaction of this phase is considerably more intense than in the $\gamma$ phase\cite{sakuma2003first}. This behavior shows itself in the Néel temperature of the two materials, which is around 960 K for L1\textsubscript{2}-IrMn\textsubscript{3}\cite{doi:10.1063/1.371298}, compared to 730 K for $\gamma$-IrMn\textsubscript{3}\cite{doi:10.1143/JPSJ.36.445}. Moreover, Kohn \textit{et. al.}\cite{kohn2013antiferromagnetic}, have experimentally measured hysteresis loops for L1\textsubscript{2}-IrMn\textsubscript{3}/Fe, and $\gamma$-IrMn\textsubscript{3}/Fe, discovering a larger exchange bias field and coercitivity for the L1\textsubscript{2} phase, compared to the $\gamma$ phase.

\begin{figure}[htbp]
    \centering
    \includegraphics[width=10cm]{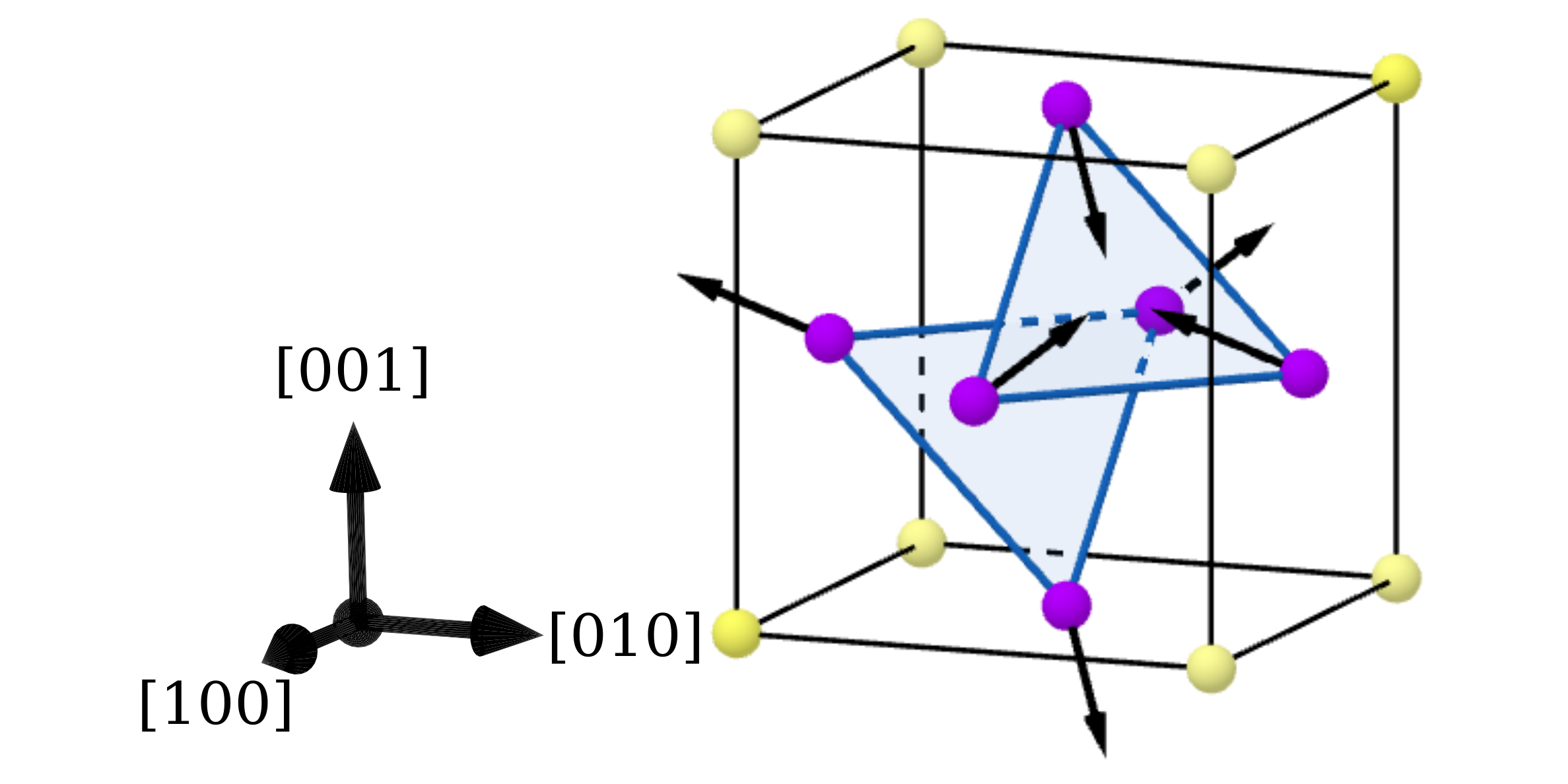}
    \caption{Crystal structure of the L1\textsubscript{2}-IrMn\textsubscript{3} phase with a T1 non-collinear antiferromagnetic structure. The iridium and manganese atoms are represented in yellow and purple, respectively.}
    \label{fig:main_1}
\end{figure}

L1\textsubscript{2}-IrMn\textsubscript{3} is a non-collinear antiferromagnetic material, with a triangular T1 magnetic structure. Its atoms are arranged in a face-centered cubic (FCC) lattice, corresponding to space group 221, and with a room-temperature lattice parameter $a=3.77$ \AA\cite{doi:10.1063/1.371298}. Regarding the T1 magnetic structure, the magnetic moments of the Mn atoms are parallel to the \{111\} planes, and they are aligned in the <112> directions, as shown in Fig. \ref{fig:main_1}. This magnetic arrangement has been obtained computationally\cite{sakuma2003first}, and experimentally, using neutron dispersion\cite{doi:10.1063/1.371298}. 

IrMn\textsubscript{3} has been studied in bulk and in heterostructures, both experimentally\cite{kohn2013antiferromagnetic,Zhange1600759,SHARMA2019165390,JaraArticle} and theoretically\cite{sakuma2003first,8f1735622d0c42f99d531bb0b0069f94,articleSzunyogh2011}. However, to the best of our knowledge, there are no studies of the thermodynamic surface stability or the magnetic surface behavior of the L1\textsubscript{2} phase of IrMn\textsubscript{3}. These are fundamental physical aspects, which must be described, to fully understand its properties in heterostructures, and to exploit them in technical applications. In the present work, we study the L1\textsubscript{2}-IrMn\textsubscript{3} surface, using first-principles Density Functional Theory computations, predicting and describing its most favorable atomic and magnetic configuration. Furthermore, we study the energetic stability of the L1\textsubscript{2}-IrMn\textsubscript{3}/Fe heterostructure. In particular, we describe the atomic arrangement and magnetic coupling at the interface and discuss their relationship with the EB effect. Kohn. \textit{et. al.}\cite{kohn2013antiferromagnetic} have previously performed experimental research on the IrMn\textsubscript{3}/Fe heterostructure. However, the magnetic coupling for individual atoms at the interface is still unknown, and it could provide valuable information in terms of understanding and engineering the exchange bias properties of this system. Our aim with this study is to provide a detailed analysis of the thermodynamic stability, structural and magnetic properties of the L1\textsubscript{2}-IrMn\textsubscript{3} surface and the L1\textsubscript{2}-IrMn\textsubscript{3}/Fe heterostructure. Thus, 
enhancing our understanding of their atomic structure, non-trivial surface and interface magnetic arrangements, and exchange-bias properties.

Our paper is organized as follows: in the results section, we propose several possible configurations of the IrMn\textsubscript{3} surface, and compare their thermodynamic stability. We find two stable surface configurations that are consistent with experimental results. Moreover, we provide a description of the most stable atomic and magnetic surface arrangements. Then, we use these findings to generate four different configurations of the IrMn\textsubscript{3}/Fe heterostructure and compare them using a thermodynamic analysis. After obtaining the most stable heterostructures, we describe the atomic and magnetic arrangements at the interface. Furthermore, we investigate the exchange-bias properties of the system using a comparative approach. We discover a relationship between exchange-bias intensity and Mn-Fe interactions at the interface, which could lead to EB tailoring by selective growth of the heterostructure. Moreover, we find a relationship between AF order and interface stability, which provides a possible mechanism for the appearance of EB in non-collinear/collinear heterostructures. Our calculations are in good agreement with experimental results on this bilayer system. Finally, in the last section, we make conclusions.

\section{Results and Discussion}

\subsection{Bulk IrMn\textsubscript{3}}

To confirm the stability of the T1 magnetic structure we have fully optimized the IrMn\textsubscript{3} bulk structure in the following magnetic configurations: T1-antiferromagnetic (T1-AF), ferromagnetic, ferrimagnetic (FiM), and non-magnetic (NM).  Our findings are summarized in Table \ref{tab:1}. It is important to mention that in previous studies on the IrMn\textsubscript{3}/Co heterostructre\cite{yanes2013exchange}, it has been noted that the exchange bias and magnetic coupling in this system are heavily influenced by the Dzyaloshinskii-Moriya (DM) interaction, which originates from relativistic corrections, i.e. Spin-Orbit Coupling (SOC). Therefore, to accurately describe the magnetic properties of IrMn\textsubscript{3}, SOC has been included in all bulk, surface, and heterostructure calculations.

\begin{table}[htbp]
\caption{Optimized lattice parameters and energy differences for bulk IrMn\textsubscript{3} structures.}
\label{tab:1}
\begin{tabular}{ccc}
\hline
\multicolumn{1}{l}{Magnetic Arrangement} & \multicolumn{1}{l}{Lattice parameter (\AA)} & \multicolumn{1}{l}{$\Delta E$ (eV)} \\ \hline
T1-AF                                    & 3.70                                                       & 0.00                                \\
FiM                                      & 3.68                                                       & 0.15                                \\
FM                                       & 3.61                                                       & 0.73                                \\
NM                                       & 3.60                                                       & 1.17                                \\
Exp.                                     & 3.77                                                       &                                     \\ \hline
\end{tabular}
\end{table}

As expected, the triangular T1 antiferromagnetic configuration is the most stable, while the FiM, FM and NM structures have higher energies of 0.15 eV, 0.73 eV and 1.17 eV, respectively. For the T1 configuration, the magnetic and crystal structures are maintained after relaxation. Our optimized lattice parameter, $a=3.70$ \AA~, is in good agreement with the room-temperature experimental value of 3.77 \AA\cite{doi:10.1063/1.371298}. Aditionally, our calculated Mn magnetic moments have a magnitude of 2.66 $\mu_B$, in good agreement with reported values of 2.62 $\mu_B$\cite{sakuma2003first}. We then use the relaxed parameters as a starting point for the following surface and heterostructure calculations.

\subsection{IrMn\textsubscript{3} Surface}

\subsubsection*{Thermodynamic Stability}

Due to symmetry, the ground state of the T1 magnetic configuration in bulk IrMn\textsubscript{3} has eight equivalent magnetic configurations. However, this symmetry is broken at the surface, conducting to four different T1 magnetic configurations (we label them as b1, b2, b3 and b4), with different magnetic moments at the surface (see Fig. S1 in supplementary information). Due to the complicated nature of the T1 alignment, the different magnetic configurations at the surface have an impact on the stability of this system. Therefore, to find the ground state of the IrMn\textsubscript{3} surface, we must find the magnetic arrangement which gives the highest stability. Furthermore, for the atomic arrangement, we have considered two pristine surfaces (IrMn and MnMn terminating layers), and two surfaces with defects (Ir/Fe substitution and Ir vacancy at the surface). The pristine surfaces have been analyzed with all four T1 magnetic configurations, and the defect structures have been studied using only the b2 magnetic arrangement. Thus, a total of 10 proposed structures have been compared. The atomic arrangements and magnetic configurations are shown in Figure S1 of the supplementary information.

\begin{figure}[ht]
    \centering
    \includegraphics[width=16cm]{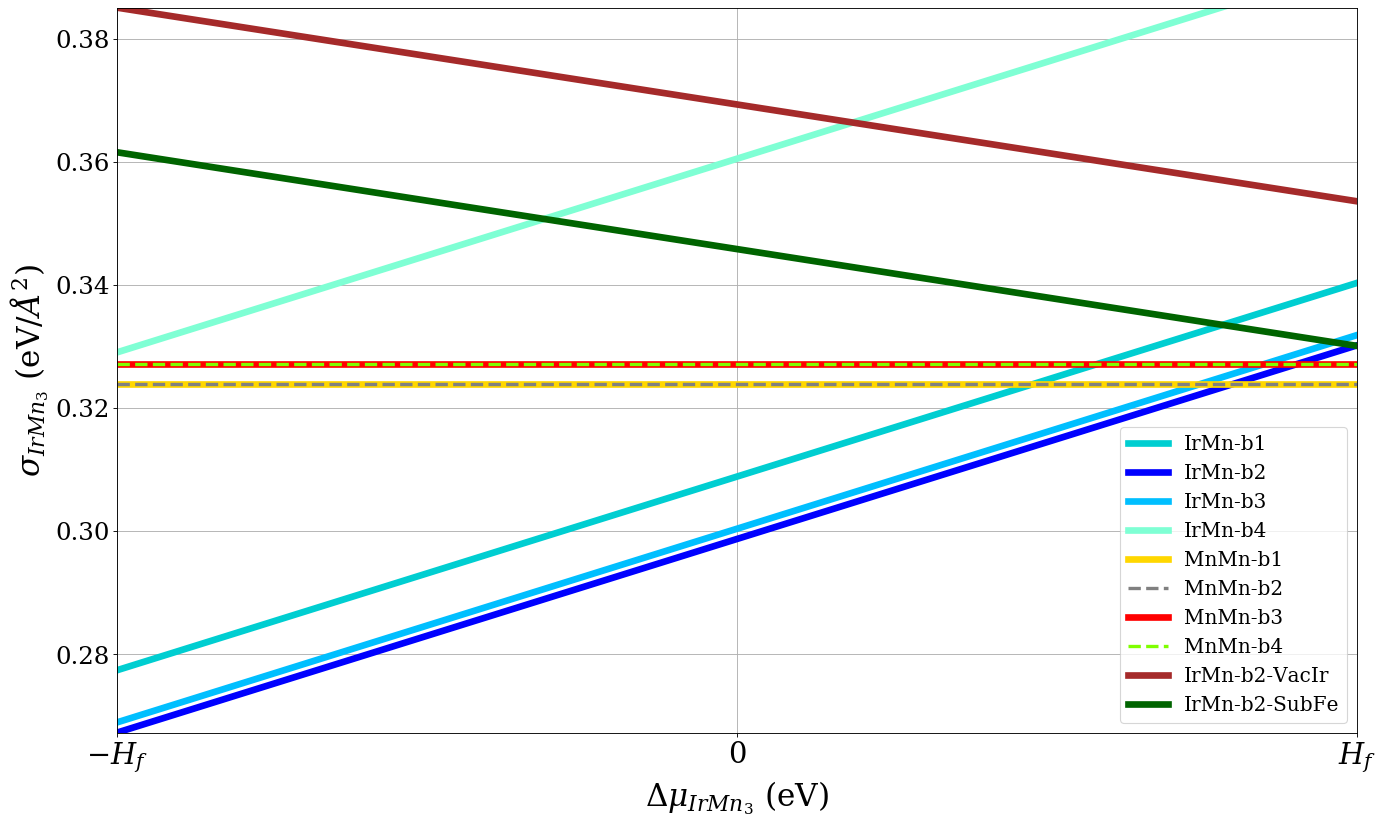}
    \caption{L1\textsubscript{2}-IrMn\textsubscript{3} surface formation energy ($\sigma_{IrMn\textsubscript{3}}$) as a function of the difference in chemical potential. The x-axis is centered at $3\mu_{Mn}-\mu_{Ir}$, such that the Ir-rich/Mn-poor conditions ($\Delta\mu_{IrMn\textsubscript{3}} = 3\mu_{Mn}-\mu_{Ir}-\Delta H_f$) are at the left side of the diagram, and the Mn-rich/Ir-poor conditions ($\Delta\mu_{IrMn\textsubscript{3}} = 3\mu_{Mn}-\mu_{Ir}+\Delta H_f$) are at the right side.}
    \label{fig:main_2}
\end{figure}

The stoichiometry of the different atomic configurations is not the same, making it impossible to directly compare their total energies. Thus, we have analyzed the stability of our 10 proposed structures using a surface formation energy (SFE) formalism described by Guerrero-Sanchez, \textit{et. al.}\cite{guerrero2015structural}, and adapted for the IrMn\textsubscript{3} system. This formalism allows us to graphically compare the SFE of each configuration using a stability diagram, which depends on the chemical potential of the species involved. The stability diagram is presented in Fig. \ref{fig:main_2}. The variable $\Delta \mu_{IrMn\textsubscript{3}}$ is defined as $3\mu_{Mn}-\mu_{Ir}$, and the limits on the x-axis correspond to Ir-rich/Mn-poor ($-\Delta H_f$) and Ir-poor/Mn-rich ($\Delta H_f$) growth conditions. Our results show that there are three stable surface structures for this system: IrMn-b2, MnMn-b1, and MnMn-b2. Structures MnMn-b1 and MnMn-b2 present the same energy and are, in fact, symmetrically equivalent structures. Therefore, we will refer to them as the MnMn-b1/b2 structure. On the other hand, terminations IrMn-b1 and IrMn-b2 are not equivalent and, although the difference in energy between the two is small, only one of them (IrMn-b2) is stable. These results indicate that the L1\textsubscript{2}-IrMn\textsubscript{3} surface structure has a preference to arrange its superficial magnetic moments in the b1 and b2 magnetic configurations.

The stability diagram shows favorable growth of the IrMn-b2 termination for a wide range of chemical potentials, covering almost all growth conditions, except at the Ir-poor/Mn-rich limit. On the other hand, we see favorable growth for the MnMn-b1/b2 termination in a limited range of chemical potentials near the Ir-poor/Mn-rich limit. None of the defect surface structures are stable. Therefore, our results predict that there are two ideal L1\textsubscript{2}-IrMn\textsubscript{3} terminations that provide high stability for the system. Moreover, the synthesis of either structure could be tuned by modifying the growth conditions. Experimentally, it has been shown\cite{kohn2013antiferromagnetic} that the layer termination of IrMn\textsubscript{3} alternate between IrMn and MnMn, when synthesizing epitaxial layers of IrMn\textsubscript{3} coupled with BCC iron.

\subsubsection*{Structural and Magnetic Properties}

\begin{figure}[htbp]
    \centering
    \includegraphics[width=16cm]{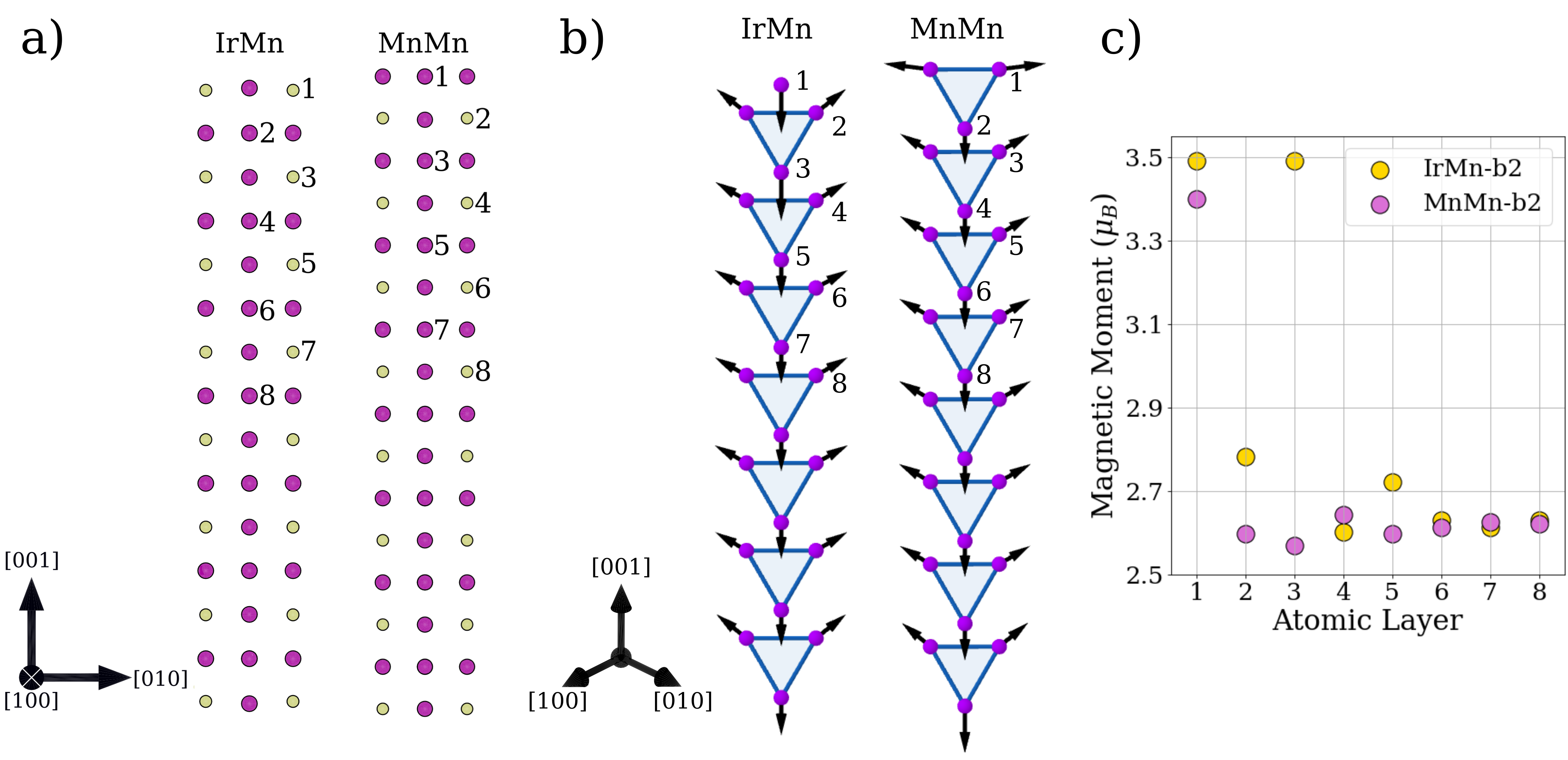}
    \caption{Optimized L1\textsubscript{2}-IrMn\textsubscript{3} surface structures with IrMn-b2 and MnMn-b2 terminating layers. a) Atomic structure. b) Magnetic arrangement. The Ir and Mn atoms are represented as yellow and purple circles, respectively. Atoms of interest are labeled 1-8. c) Magnitude of the magnetic moments near the surface for IrMn and MnMn terminations.}%
    \label{fig:main_3}
\end{figure}

We now describe the structural and magnetic properties of the optimized IrMn-b2 and MnMn-b1/b2 terminations of the L1\textsubscript{2}-IrMn\textsubscript{3} surface. Fig. \ref{fig:main_3}a shows the optimized surface structures, while the interatomic distances are summarized in Table \ref{tab:2}. The reference atoms in Table \ref{tab:2} are labeled numerically in Fig. \ref{fig:main_3}a. For the IrMn termination, we find a slight rearrangement in the atomic structure near the surface, which extends through the first four atomic layers. The first two layers suffer a compression, with distances of 2.59 \AA~ between the Ir and Mn atoms, while layers 2-4 are elongated with Ir-Mn distances of 2.63 \AA~ and 2.62 \AA. For layers 4-8 the bulk structure is preserved, with Ir-Mn distances near the bulk value of 2.61 \AA. Similarly, for the MnMn terminated surface, the first two layers suffer compression and layers 2-3 are elongated. The following six layers preserve the bulk atomic arrangement. The modified structural parameters near the surface are attributed to the lower coordination number (less neighboring atoms) in superficial atoms, which alters the interatomic forces at the surface and leads to a distortion in the atomic spacing of terminating layers.

\begin{table}[htbp]
    \centering
    \begin{tabular}{ccc}
    \hline
    \multirow{2}{*}{\textbf{Bond}} & \multicolumn{2}{c}{\textbf{Interatomic Distance (\AA)}} \\
                                 & \textit{IrMn}                     & \textit{MnMn}                    \\ \hline
1-2                              &          2.59                  & 2.60                             \\
2-3                              &          2.62                  & 2.63                             \\
3-4                              &          2.63                  & 2.61                             \\
4-5                              &          2.61                  & 2.62                             \\
5-6                              &          2.62                  & 2.61                             \\
6-7                              &          2.61                  & 2.61                             \\
7-8                              &          2.62                  & 2.61                             \\ \hline
    \end{tabular}
    \caption{Interatomic distances in L1\textsubscript{2}-IrMn\textsubscript{3}, with IrMn-b2 and MnMn-b2 terminating layers.}
    \label{tab:2}
\end{table}
\vspace{-10pt}

To better understand the magnetic behavior of the L1\textsubscript{2}-IrMn\textsubscript{3} system at the surface, we have plotted the magnetic moments of the Mn atoms (Fig. \ref{fig:main_3}b) and their magnitude close to the surface (Fig. \ref{fig:main_3}c) for the two stable configurations. In both terminations, the T1 magnetic configuration is preserved in the interior of the slabs. However, the spin arrangement at the surface is distorted. For the IrMn terminating layer, we see an increase in the magnitude of the magnetic moments near the surface (see Fig. \ref{fig:main_3}c). Furthermore, there is a slight tilt ($\sim 1.2^\circ$) in the alignment of atom 1 with respect to the atoms in the interior region of the structure. On the other hand, the magnetic moments of the MnMn termination present an increase in magnitude (see Fig. \ref{fig:main_3}c), and a strong reorientation at the surface. In this case, the magnetic alignment at the terminating layer presents a decrease in the $c$ [001] component and an increase in the $a$ [100] and $b$ [010] components in comparison with interior atoms. The angular tilt between the magnetic moments of atoms 1 and 7 is $\sim23^\circ$. Values of the magnetic moments for atoms 1-8 in Fig. \ref{fig:main_3}b are supplied in Table S1 of the supplementary information.

As a general trend, we can say that the surface atoms present a rearrangement, characterized by a compression in the outermost layer and an elongation in the inner layer(s). Also, the superficial magnetic moments show an increase in their magnitude and a change in their direction. These surface phenomena are attributed to the lower coordination number and, thus, modified interatomic forces at surface atoms, compared to the bulk-like atoms in the interior region.

\subsection{IrMn\textsubscript{3}/Fe Heterostructure}
The exchange-coupled L1\textsubscript{2}-IrMn\textsubscript{3}/Fe bilayer has previously been experimentally built using molecular beam epitaxy, and its exchange-bias properties have been investigated using vibrating sample magnetometry.\cite{kohn2013antiferromagnetic} However, it is currently not possible to experimentally determine the interface magnetic coupling, atom-by-atom. We have, first, obtained the most stable heterostructures, and then, we have described and analyzed the atomic arrangement and magnetic coupling of each atom at the interface. We have not only replicated the experimental findings, but we have also been able to determine general trends in the system which lead to a better description of exchange bias in non-collinear/collinear bilayer systems.

\subsubsection*{Thermodynamic Stability}
\begin{figure}[htbp]
    \centering
    \includegraphics[width=16.5cm]{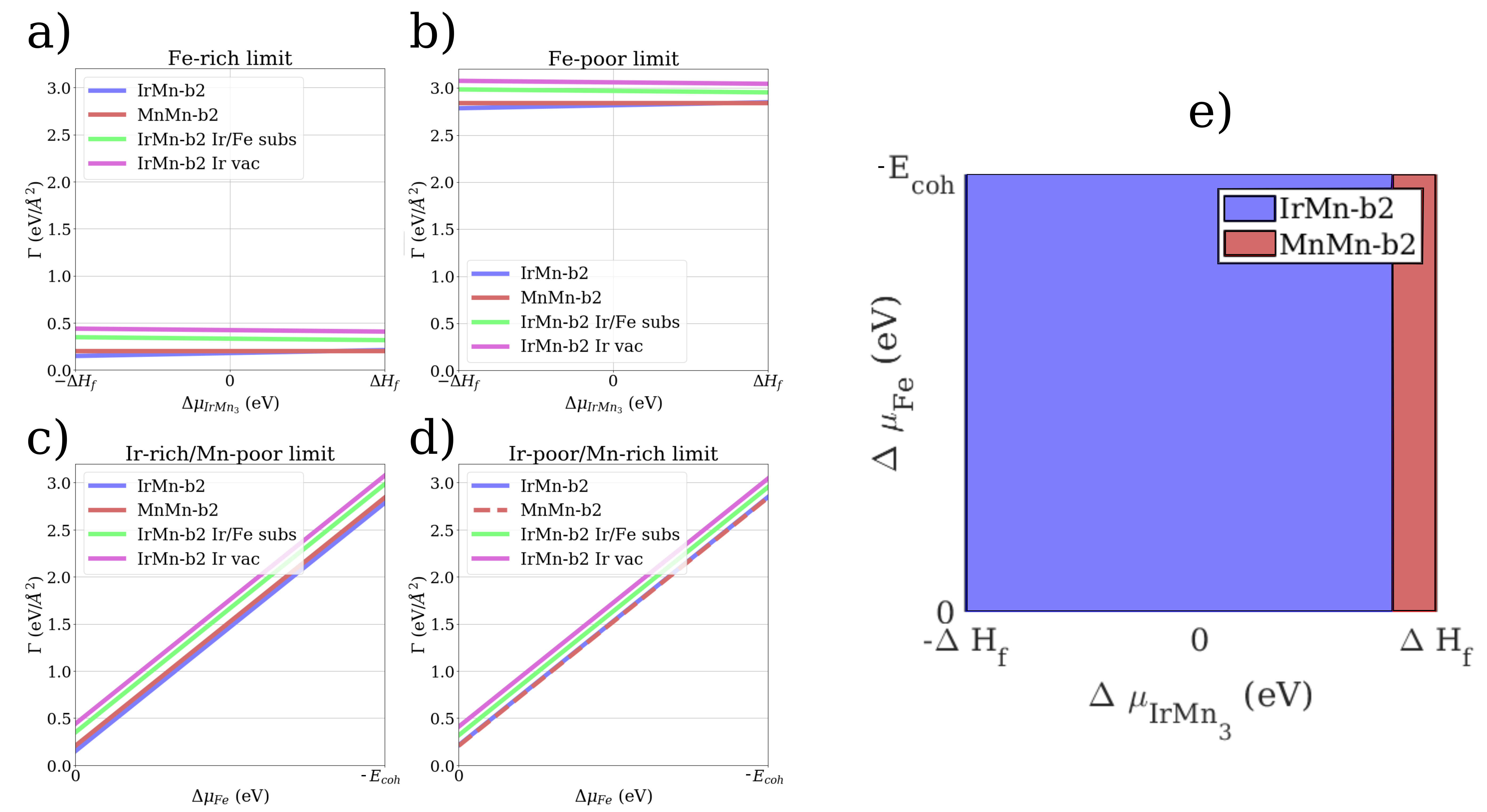}
    \caption{Interface formation energy plot for the IrMn3/Fe system. a-d) 2D projections of the 3D planes at different chemical potential limits. e) Projection on the x-y axis (stability diagram), showing the two structures with highest stability.}%
    \label{fig:main_4}
\end{figure}

In the previous section, we have obtained and characterized the energetically favorable surface terminations for L1\textsubscript{2}-IrMn\textsubscript{3}. In this subsection, we have built a heterostructure by coupling the optimized IrMn\textsubscript{3} surfaces with a 7 layer body-centered cubic (BCC) Fe(001) surface, magnetized in the Fe[00-1] direction. We have, then, fully optimized the atomic positions and magnetic moments of four heterostructures, with different layer terminations: IrMn, MnMn, Ir-vacancy, and Fe-substitution. 
The four generated heterostructures have different stoichiometries. Therefore, their stability is compared using an interface formation energy formalism following the work of Liu, \textit{et. al}\cite{liu2005first}, adapted to the IrMn\textsubscript{3}/Fe system (see Methods section). The interface formation energy is plotted as a bi-linear function of the chemical potential differences for IrMn\textsubscript{3} and Fe, resulting in three-dimensional planes that cover all of the allowed chemical potentials (see Figure S2 in supplementary information for the 3D plot). In the $\Delta\mu_{IrMn\textsubscript{3}}$ axis, $-\Delta H_f$ and $\Delta H_f$ correspond to Ir-rich/Mn-poor and Ir-poor/Mn-rich growth conditions, respectively. The $\Delta\mu_{Fe}$ axis spans from Fe-rich(0) to Fe-poor($-E_{coh}$) growth conditions. In order to better visualize the stability planes, we plot 2D projections of the three-dimensional plot, as shown in Fig. \ref{fig:main_4}. 

Figs. \ref{fig:main_4}a, and \ref{fig:main_4}b depict the interface formation energy lines for Fe-rich and Fe-poor growth conditions, respectively. Here, we see that for most of the chemical potential range, the IrMn-b2 system has higher stability. The exception is near the Mn-rich chemical potential difference, where the MnMn-b2 terminated system is more stable. Moreover, we see that the energy difference between the stable terminations is small for all the range of chemical potentials, indicating that it is likely to have both, IrMn, and MnMn interface terminations. This result explains the findings of Kohn, \textit{et. al.}\cite{kohn2013antiferromagnetic}, who synthesized high quality epitaxial L1\textsubscript{2}-IrMn\textsubscript{3}/Fe bilayers and consistently found chemically sharp IrMn and MnMn terminating layers, which alternate within the same sample and with no control on the interface formation.

Figs. \ref{fig:main_4}c and \ref{fig:main_4}d show the interface formation energy lines for Ir-rich/Mn-poor and Ir-poor/Mn-rich growth conditions, respectively. We see that the interfaces present the highest stability at Fe-rich conditions ($\Delta \mu_{Fe} = 0$). From these figures, it is apparent that the Fe-substitution and Ir-vacancy interface structures are the least favorable for the IrMn\textsubscript{3}/Fe heterostructure. These results are in good agreement with the experimental work of Kohn, \textit{et. al.}\cite{kohn2013antiferromagnetic}, who observed negligible chemical intermixing between the IrMn\textsubscript{3} and the Fe coupled layers. This is explained by the relatively low stability of the Fe-substitution terminated layer, represented by a green line in the chemical potential limit projections.

Figure \ref{fig:main_4}e is a projection of the 3D planes onto the x-y axis. Although we have plotted 3D stability planes for four different structures, this stability diagram allows us to exclusively visualize the most stable ones. With this diagram, we can analyze the growth conditions needed to synthesize different terminations and the ranges where they could occur. We see that IrMn and MnMn are the only interface terminations that minimize the formation energy. The IrMn interface is stable in a wide spread of chemical potentials, including Ir-rich/Mn-poor and Ir/Mn moderate growth conditions, and throughout the Fe growth conditions. Meanwhile, the MnMn interface presents higher stability in a narrow range of chemical potentials, near the Ir-poor/Mn-rich limit. Thus, Ir-rich conditions are more likely to result in the formation of IrMn terminations, while Mn-rich conditions are more likely to result in  MnMn terminations. This selective growth can be used to control the layer termination, which allows access to specific properties for this material.

\subsubsection*{Structural Properties}

\begin{figure}[htbp]
    \centering
    \includegraphics[width=16.5cm]{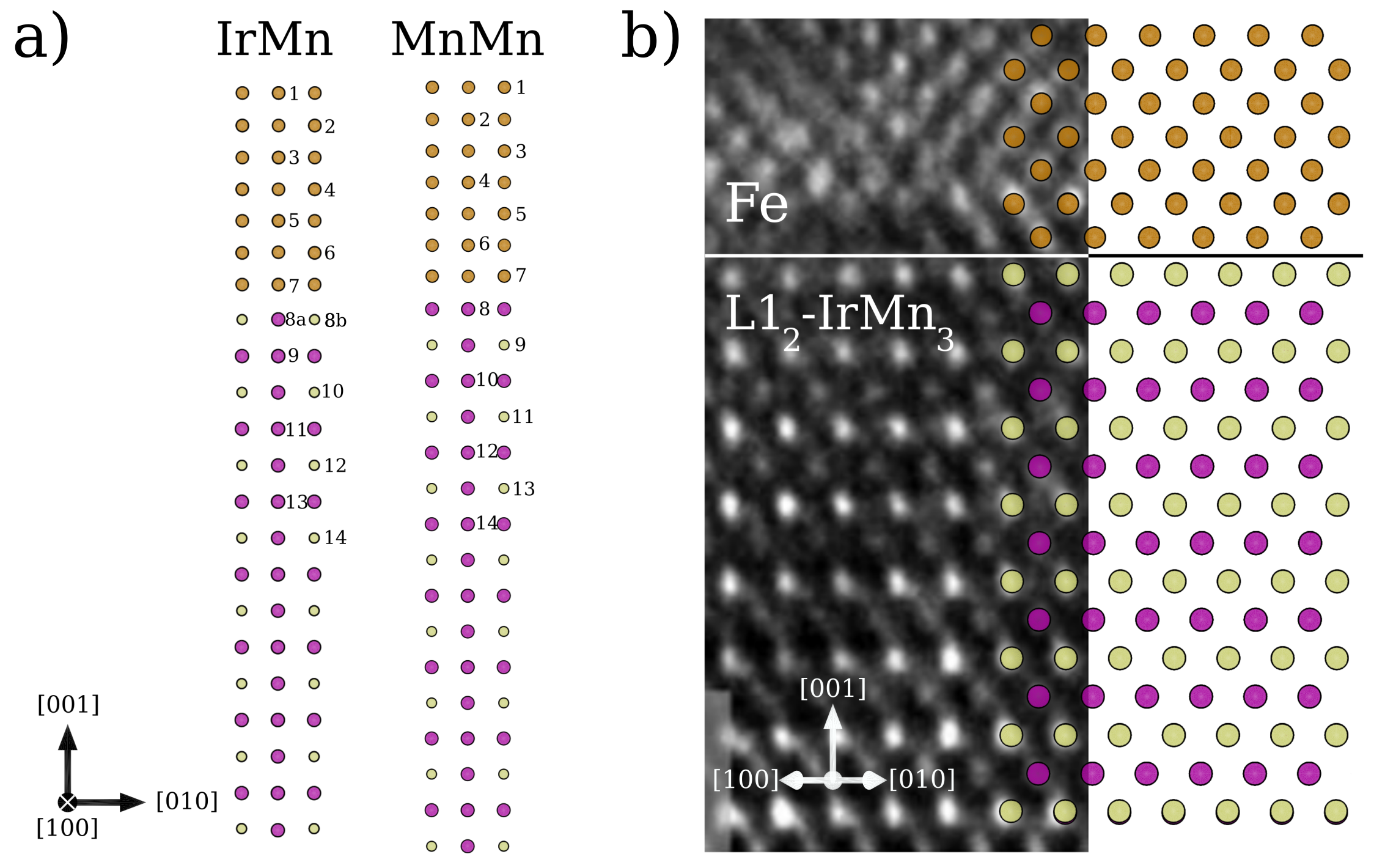}
    \caption{Stable models of the IrMn\textsubscript{3}/Fe heterostructure. a) IrMn and MnMn terminated layer structures. The atoms of interest are labeled 1-14 in each structure. b) Comparison between the computational optimized IrMn terminated structure and the HR-TEM data. The experimental image was provided by Kohn, \textit{et. al.}\cite{kohn2013antiferromagnetic}. The axes represent the [100], [010] and [001] crystallographic directions for IrMn\textsubscript{3}, which are equivalent to the Fe[1-10], Fe[110] and Fe[001] crystallographic directions.}%
    \label{fig:main_5}
\end{figure}

The two stable structures at different growth conditions, found using the interface formation analysis, are shown in Figure \ref{fig:main_5}a. The structural parameters were obtained after full optimization of the atomic positions and magnetic moments. Selected bond distances are presented in Table \ref{tab:3}.

For the IrMn termination, we observe a small elongation for all Fe interlayers: 2.48 \AA~ between layers 1-2, 2.46 \AA~ between layers 2-3, followed by an interatomic distance of 2.45 \AA~ in the interior of the Fe slab, and a slightly elongated distance of 2.46 \AA~ in layers adjacent to the interface (the optimized bulk distance is 2.40 \AA). However, these distances are similar to the ones of the optimized Fe surface before being coupled to IrMn\textsubscript{3}, which shows bond distances of 2.44-2.45 \AA~ in the interior region of the slab. At the IrMn\textsubscript{3}/Fe interface there are two types of bonds: Fe-Mn, and Fe-Ir (corresponding to distances 7-8a and 7-8b, respectively). The interface bonds have a distance which falls between the Fe-Fe and Ir-Mn distances, allowing for a structurally smooth transition from Fe to IrMn\textsubscript{3}. The following atoms (8-14) have interatomic distances which preserve the bulk structure of L1\textsubscript{2}-IrMn\textsubscript{3}, yet are slightly elongated by 0.01 \AA. 

Similarly, the interlayer distance between the first Fe layers (1-2) of the MnMn terminated surface presents an elongation, followed by an interatomic distance of 2.45 \AA~ in the interior region of the slab. In this case, there is higher disorder near the interface, with the Fe-Fe bond distances elongating (2.46 \AA) between layers 5-6, and contracting (2.43 \AA) between layers 6-7. At the interface, the Fe-Mn distance is 2.50 \AA. Again, this bond distance falls between the Fe-Fe and Ir-Mn bond distances, providing a smooth transition between the two slabs. Finally, the interatomic distances in the interior region of the IrMn\textsubscript{3} slab are consistent with the bulk structure, with some of the distances being slightly contracted by 0.01 \AA.

\begin{table}[htbp]
\caption{Optimized lattice distances for the two stable IrMn\textsubscript{3}/Fe interface terminations.}
\label{tab:3}
\begin{tabular}{cccc}
\hline
\multicolumn{2}{l}{IrMn-b2 termination}                            & \multicolumn{2}{|l}{MnMn-b2 termination}                                              \\
Bonding Atoms & \multicolumn{1}{l|}{Distance (\AA)} & \multicolumn{1}{l}{Bonding atoms} & \multicolumn{1}{l}{Distance (\AA)} \\ \hline
1-2           & 2.48                                               & 1-2                               & 2.48                                              \\
2-3           & 2.46                                               & 2-3                               & 2.45                                              \\
3-4           & 2.45                                               & 3-4                               & 2.45                                              \\
4-5           & 2.45                                               & 4-5                               & 2.44                                              \\
5-6           & 2.46                                               & 5-6                               & 2.46                                              \\
6-7           & 2.46                                               & 6-7                               & 2.43                                              \\
7-8a          & 2.56                                               & \multirow{2}{*}{7-8}              & \multirow{2}{*}{2.50}                             \\
7-8b          & 2.57                                               &                                   &                                                   \\
8-9           & 2.62                                               & 8-9                               & 2.60                                              \\
9-10          & 2.61                                               & 9-10                              & 2.61                                              \\
10-11         & 2.62                                               & 10-11                             & 2.60                                              \\
11-12         & 2.62                                               & 11-12                             & 2.61                                              \\
12-13         & 2.62                                               & 12-13                             & 2.60                                              \\
13-14         & 2.62                                               & 13-14                             & 2.61                                              \\ \hline
\end{tabular}
\end{table}

An interesting observation is that near the interface, there is an exact match (2.61 \AA) in the in-plane (a-b plane) optimized interatomic distances of IrMn\textsubscript{3} and Fe. This allows for a smooth transition between the two structures, further contributing to the high stability of the IrMn and MnMn terminations. The key-role of having a good match in lattice parameters becomes evident from Fig. \ref{fig:main_5}b, which shows a comparison between our optimized computational structure and a high-resolution TEM image of the L1\textsubscript{2}-IrMn\textsubscript{3}/Fe bilayer. Our computations successfully reproduced the experimental data, showing a uniform transition between the IrMn\textsubscript{3} and Fe surfaces, and an excellent match for the in-plane lattice parameters at the interface.

\subsubsection*{Magnetic Properties}

In this final subsection, we analyze the magnetic properties of the IrMn\textsubscript{3}/Fe system. First, we investigate the exchange-bias properties of the two most stable configurations. Following the work of Dong, et. al.\cite{dong2011ab}, we employ a comparative approach to analyze the exchange bias effect in our system. Using the optimized structures as a template, we calculate the total energies for different magnetizations of the heterostructure, by self-consistent calculations (SCF). We set the IrMn\textsubscript{3} slab in the b2 magnetic configuration, which was maintained for all SCF calculations. First, we calculate three magnetic arrangements, where the Fe slab was initially magnetized in the Fe[100], Fe[001] and Fe[110] directions. We then flip the magnetization of the Fe slab to the Fe[-100], Fe[00-1] and Fe[-1-10] directions, and ran SCF calculations using the same computational parameters as the previous ones. The total energies were analyzed by comparing spin-reversed pairs, as shown in Table \ref{tab:4}. The energy difference ($\Delta E$) between the pairs allows us to investigate the intrinsic exchange-bias present in the system: larger $\Delta E$ indicates a higher bias effect for that configuration. 

\begin{table}[htbp]
\caption{Total energies from SCF calculations and energy difference for Fe spin-inverted pairs of the IrMn\textsubscript{3}/Fe heterostructure.}
\label{tab:4}
\begin{tabular}{ccccc}
\hline
\multirow{2}{*}{\begin{tabular}[c]{@{}c@{}}Magnetic Moment \\ Directions\end{tabular}} & \multicolumn{2}{c}{IrMn termination}                   & \multicolumn{2}{c}{MnMn termination}                   \\
                                                                                       & \textit{Total Energy (eV)} & \textit{$\Delta E$ (meV)} & \textit{Total Energy (eV)} & \textit{$\Delta E$ (meV)} \\ \hline
Fe{[}100{]}                                                                            & -390.5321                  & \multirow{2}{*}{29.2}     & -408.5506                  & \multirow{2}{*}{93.2}     \\
Fe{[}-100{]}                                                                           & -390.5613                  &                           & -408.6438                  &                           \\
Fe{[}001{]}                                                                            & -390.5767                  & \multirow{2}{*}{18.5}     & -408.6998                  & \multirow{2}{*}{85.0}     \\
Fe{[}00-1{]}                                                                           & -390.5952                  &                           & -408.6148                  &                           \\
Fe{[}110{]}                                                                            & -390.5356                  & \multirow{2}{*}{17.9}     & -408.5823                  & \multirow{2}{*}{72.7}     \\
Fe{[}-1-10{]}                                                                          & -390.5535                  &                           & -408.6550                  &                           \\ \hline
\end{tabular}
\end{table}

In good agreement with experimental hysteresis measurements of the IrMn\textsubscript{3}/Fe system\cite{kohn2013antiferromagnetic}, our results indicate that exchange bias is enhanced when the  magnetization is along the Fe[100] direction (Fe easy-axis), compared to the Fe[110] direction (Fe minor hard-axis). Interestingly, an intermediate value arises when the magnetization is in the Fe[001] direction, which has not yet been experimentally investigated. From our results in Table \ref{tab:4}, we discover that the MnMn termination presents larger energy differences and, thus, larger exchange bias than the IrMn termination. Previous studies by Szunyogh, \textit{et. al.}\cite{articleSzunyogh2011} show that the magnetic anisotropy for the IrMn\textsubscript{3}/Co heterostructure is heavily influenced by Dzyaloshinskii-Moriya interactions at the interface. Also, Yanes, \textit{et. al}\cite{yanes2013exchange} showed that DM interations between Mn and Co are a key contribution to the exchange bias in this system. Furthermore, previous studies\cite{jadaun2020microscopic} show that the Ir/Fe DM interaction is relatively small (-1.9 meV), especially when compared with compounds that contain Mn-Fe interactions, for example Mn\textsubscript{1-x}Fe\textsubscript{x}Ge\cite{gayles2015dzyaloshinskii} (up to 10 meV).  As previously mentioned, in our heterostructure we observe larger exchange bias for the MnMn structure, which has more Mn atoms in the immediate vicinity of the Fe slab than the IrMn termination. Therefore, there is a direct relationship between the amount of Mn-Fe interactions and the exchange bias intensity. This finding could lead to EB tailoring by controlling the structure of the interface, without the need to introduce defects\cite{sort2005tailoring}, change the layer thickness\cite{sort2005tailoring,baltz2005tailoring,zhou2017robust}, or modify the morphology of the sample\cite{liu2001tailoring}.

\begin{figure}[htbp]
    \centering
    \includegraphics[width=16.5cm]{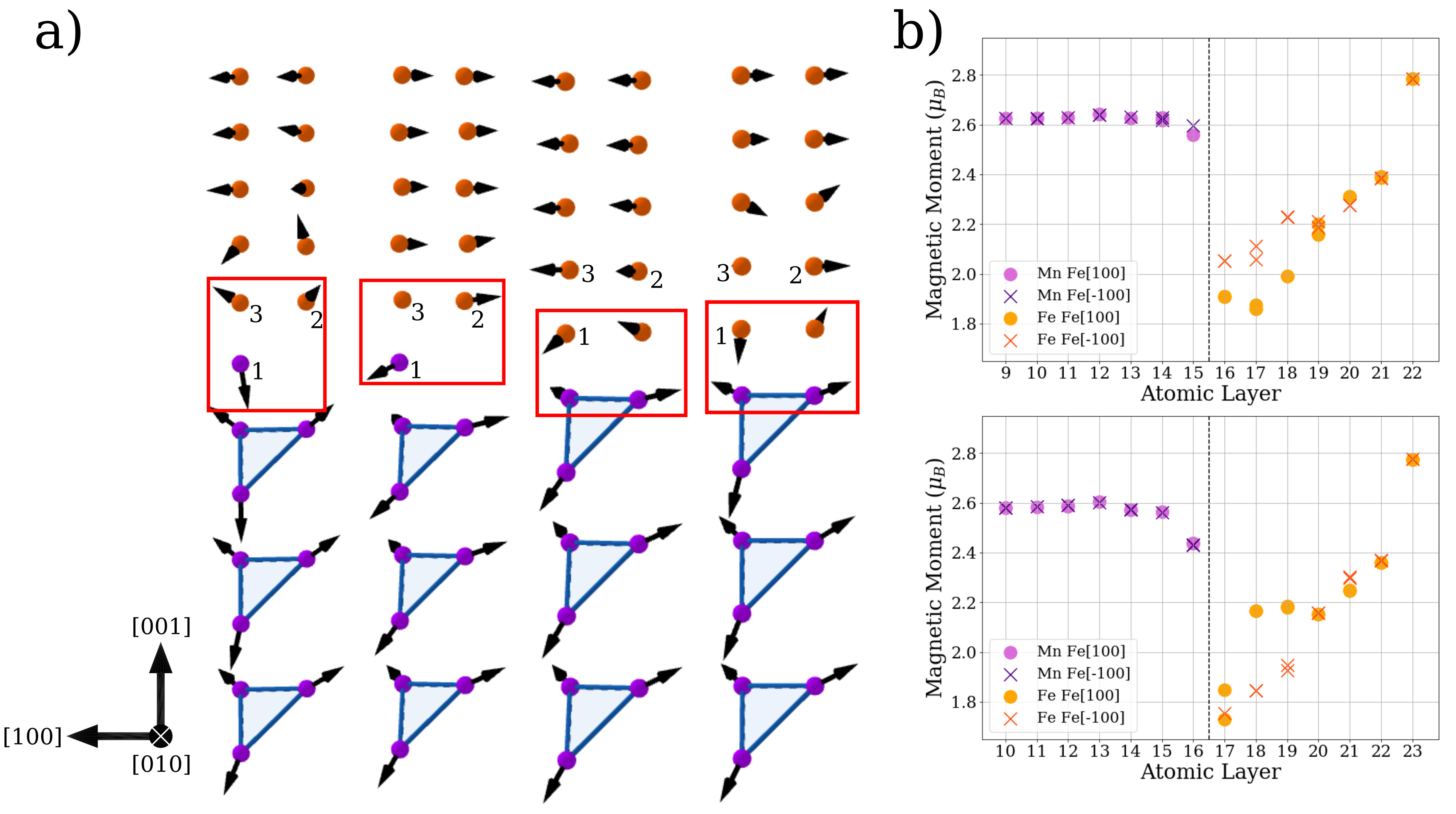}
    \caption{Magnetic coupling behavior of the L1\textsubscript{2}-IrMn\textsubscript{3}/Fe interface, with the Fe layer initially magnetized in the Fe[100] and Fe[-100] directions. a) From left to right: IrMn-b2 termination with initial magnetization in the Fe[100] direction, IrMn-b2 termination with  initial magnetization in the Fe[-100] direction , MnMn-b2 termination with  initial magnetization in the Fe[100], and MnMn-b2 termination with initial magnetization in the  Fe[-100] direction. The axes represent the [100], [010], and [001] crystallographic directions for IrMn\textsubscript{3}, which are equivalent to the Fe[1-10], Fe[110] and Fe[001] crystallographic directions. Atoms in the immediate vicinity of the interface are marked by a red rectangle. b) Magnetic moments for the IrMn (top) and MnMn (bottom) terminations at each atomic layer. The interface is marked with a dotted line.}%
    \label{fig:main_6}
\end{figure}

To better comprehend the complex magnetic coupling of these interface structures, we plot the Mn and Fe atoms near the interface with their corresponding magnetic moments. We also plot the magnitude of the magnetic moments near the interface. As an example, we show the IrMn and MnMn terminations with the Fe slab initially magnetized in the Fe[100] and Fe[-100] directions (see Fig. \ref{fig:main_6}). The magnetic coupling data for other Fe magnetization directions is provided in Figures S3-S4 of the supplementary information.

From Figure \ref{fig:main_6}a, we observe that the magnetic coupling near the interface has little influence in the IrMn\textsubscript{3} T1-AF configuration. However, the first Fe layers near the interface are heavily modified. This modification can also be observed in the magnitude of the magnetic moments near the interface (Fig. \ref{fig:main_6}b). We find that the magnitude of the  IrMn\textsubscript{3} magnetic moment is $\sim 2.6\mu_B$ in the inner region of the slab, with a slightly lower value near the interface. On the other hand, the magnitudes of the Fe magnetic moments are heavily influenced by the IrMn\textsubscript{3} slab, and present values that range from $\sim 1.75\mu_B$, near the interface, to $\sim 2.8\mu_B$ at the surface.

The Fe magnetic moments near the interface tend to align in such a way that they oppose the IrMn\textsubscript{3} moments, maintaining the AF order. When possible, the Fe moments mimic the triangular T1 antiferromagnetic structure. To quantify this result, we measured the angular difference between the bulk T1 magnetic structure and the interface magnetic coupling. For the IrMn termination, we select one Mn and two Fe atoms at the interface. On the other hand, for the MnMn termination we select three Fe atoms directly above the interface (labeled 1-3 in Figure \ref{fig:main_6}a). We then obtain the mean angular difference (MAD) by taking an average of the three angular differences for each magnetization. Lower MAD indicates a better match with the T1-AF magnetic structure. MADs for all configurations are provided in Table S2 of the supplementary information.

The mean angular differences of the Fe[100] magnetization are 0.66 and 1.27 for IrMn and MnMn terminations, respectively. Meanwhile, the MADs of the Fe[-100] magnetization are 0.52 and 0.46 for the IrMn and MnMn terminations, respectively. These values indicate that the Fe[-100] heterostructure has a better match with the bulk T1 magnetic ordering near the interface. Furthermore, from Table \ref{tab:4}, we see that the Fe[-100] magnetization presents higher stability than the Fe[100] magnetization. In fact, we find a generalized trend for this system where the AF order is directly related to the stability of the heterostructure, i.e. structures where the Fe atoms best match the T1-AF order at the interface present higher stability compared to their spin-reversed pairs.

Due to the T1 non-collinear nature of IrMn\textsubscript{3}, when magnetizing the Fe slab in opposing directions there is always an alteration of the interface AF order. This leads to a difference in energy between the two magnetizations, and one of them is more stable than the other. Therefore, this is a possible mechanism for exchange-bias in non-collinear systems. Furthermore, our findings about the relationship between AF order and stability are in good agreement with other exchange-biased heterostructures that contain Mn\cite{guerrero2018formation, corbett2019dislocation}, where it has been proved that the most stable configurations present antiferromagnetic coupling at the interface.

\section{Conclusions}

In this paper, we provide a complete analysis of the stability, structure, and magnetic arrangement of the L1\textsubscript{2}-IrMn\textsubscript{3} surface, and the L1\textsubscript{2}-IrMn\textsubscript{3}/Fe heterostructure. We find that, due to surface and interface symmetry breaking, the preferred magnetic arrangement for this system is the newly proposed T1-b2 antiferromagnetic structure. Moreover, there are two atomic arrangements that minimize the energy of the system: IrMn termination in a wide range of chemical potentials, and MnMn in a limited range of growth conditions near the Mn-rich/Ir-poor limit. Structurally, we are able to describe the surface and interface bonding distances, and we successfully explain experimental results for the interface formation. We investigate the exchange-bias properties of the IrMn and MnMn terminated heterostructures using a comparative analysis, finding larger exchange bias for the MnMn termination. This result indicates that Mn-Fe interactions at the interface play a key role in the exchange-bias properties of the heterostructure. Moreover, we predict that synthesis in Mn-rich conditions promotes selective growth of the MnMn termination, providing higher exchange bias. This result could lead to a novel method of exchange-bias tailoring mediated through control of the terminating layer. Finally, we describe the previously unknown complex magnetic coupling of the IrMn\textsubscript{3}/Fe heterostructure, discovering a relationship between stability and antiferromagnetic order. This relationship provides a possible mechanism for the appearance and enhancement of exchange bias in non-collinear systems, which is in good agreement with previous findings.

\section{Methods}

\subsection{Computational Details}
First-principles calculations were carried out within the DFT framework using the Vienna ab initio Simulation Package (VASP)\cite{PhysRevB.54.11169}. The Projector Augmented Wave (PAW) pseudopotentials\cite{blochl1994projector,kresse1999ultrasoft} were employed to describe electron-ion interactions. The exchange-correlation functional was defined through the generalized gradient approximation, parametrized by Perdew, Burke and Ernzerhof (PBE-GGA)\cite{perdew1996generalized}. A 10x10x10 gamma-centered k-point mesh was used for bulk IrMn\textsubscript{3} calculations. On the other hand, a 10x10x1 gamma-centered k-point mesh was used for surface and heterostructure calculations. For bulk and surface L1\textsubscript{2}-IrMn\textsubscript{3} calculations the cutoff energy was set to 380 eV. For IrMn\textsubscript{3}/Fe heterostructures the energy cutoff was set to 410 eV. The convergence tolerance for all calculations was set to 0.1 meV/atom. To avoid self interactions due to the periodic boundary conditions, vacuums of 20 \AA~ and 10 \AA~ (in the z-direction) were designated for surfaces and heterostructures, respectively. Finally, due to the nature of the non-collinear L1\textsubscript{2}-IrMn\textsubscript{3} system, spin orbit coupling was employed for all calculations.

\subsection{Surface Formation Energy}
The formation energy formalism used in this paper are based on the principles of thermodynamics. Formation energy is dependent on chemical potentials; therefore, it allows us to compare the relative stability of different models even though their stoichiometry is not the same. Our goal with these formalism is to compare the energy of different structures, and find the most stable.
\subsubsection*{IrMn\textsubscript{3} surface}

The number of atoms for the proposed IrMn\textsubscript{3} surfaces is different for each atomic termination (IrMn, MnMn, defect structures); thus, the total energies from DFT calculations cannot be directly compared. This leads us to define a system of equations to analyze the energetic stability of surfaces with different terminations, at varying growth conditions. Under thermodynamic stability, the surface formation energy of IrMn\textsubscript{3} ($\sigma_{IrMn\textsubscript{3}}$) can be written as:

\begin{equation} \label{eq:1}
    \sigma_{IrMn\textsubscript{3}} = \frac{1}{A}(E^{slab}_{IrMn\textsubscript{3}} - n_{Mn}\mu_{Mn} - n_{Ir}\mu_{Ir})
\end{equation}

where $n_i$ and $\mu_i$ are the number of atoms, and chemical potential of the ith species, respectively. $A$ is the cross-sectional area of the surface and $E^{slab}_{IrMn\textsubscript{3}}$ is the surface energy obtained directly from the DFT computation. We assume that the bulk and surface structures of IrMn\textsubscript{3} are in thermal equilibrium, which implies that the chemical potential of IrMn\textsubscript{3} can be written as:

\begin{equation}\label{eq:2}
    \mu_{IrMn\textsubscript{3}}^{bulk} = \mu_{Ir} + 3 \mu_{Mn} ,
\end{equation}

with ($3\mu_{Mn}<3\mu_{Mn}^{bulk}$) and ($\mu_{Ir}<\mu_{Ir}^{bulk}$) to prevent the formation of Ir and Mn bulk phases. Furthermore, we introduce a variable $\Delta\mu_{IrMn\textsubscript{3}}$ that allows us to analyze the formation energy at different growth conditions, namely Ir-rich/Mn-poor and Ir-poor/Mn-rich. We define $\Delta\mu_{IrMn\textsubscript{3}}$ such that

\begin{equation}\label{eq:3}
    \Delta\mu_{IrMn\textsubscript{3}} = 3 \mu_{Mn} - \mu_{Ir} .
\end{equation}

Through Equations \ref{eq:2} and \ref{eq:3}, we can redefine the expressions for $\mu_{Ir}$ and $\mu_{Mn}$ in terms of $\Delta\mu_{IrMn\textsubscript{3}}$

\begin{equation}\label{eq:4}
    \mu_{Ir} = \frac{1}{2}(\mu_{IrMn\textsubscript{3}}^{bulk} - \Delta\mu_{IrMn\textsubscript{3}})
\end{equation}

\begin{equation}\label{eq:5}
    \mu_{Mn} = \frac{1}{6}(\mu_{IrMn\textsubscript{3}}^{bulk} + \Delta\mu_{IrMn\textsubscript{3}}).
\end{equation}

Using these results, we rewrite Eq. \ref{eq:1} using Eqs. \ref{eq:4} and \ref{eq:5}. With further rearrangement of the terms, we get:

\begin{equation}\label{eq:6}
        \sigma_{IrMn\textsubscript{3}} =  \frac{1}{A} \{ E^{slab}_{IrMn\textsubscript{3}} - \frac{1}{2} ( \frac{n_{Mn}}{3} + n_{Ir} ) \mu_{IrMn\textsubscript{3}}^{bulk}  - \frac{1}{2} (\frac{n_{Mn}}{3} - n_{Ir}) \Delta\mu_{IrMn\textsubscript{3}}\}
\end{equation}

Note that for the defect structure with Ir/Fe substitutions, the Fe chemical potential and number of atoms must be taken into consideration. In this case, Eq. \ref{eq:6} is modified by adding the term $n_{Fe} \mu_{Fe}$:

\begin{equation}\label{eq:7}
        \sigma_{IrMn\textsubscript{3}} =  \frac{1}{A} \{ E^{slab}_{IrMn\textsubscript{3}} - \frac{1}{2} ( \frac{n_{Mn}}{3} + n_{Ir} ) \mu_{IrMn\textsubscript{3}}^{bulk}  - \frac{1}{2} (\frac{n_{Mn}}{3} - n_{Ir}) \Delta\mu_{IrMn\textsubscript{3}} - n_{Fe} \mu_{Fe} \}
\end{equation}

The parameter $\Delta\mu_{IrMn\textsubscript{3}}$ corresponds to the different growth conditions and it varies in terms of the formation enthalpy, which we define as $\Delta H_f = \mu_{Ir}^{bulk} + 3\mu_{Mn}^{bulk} - \mu_{IrMn\textsubscript{3}}^{bulk}$. The formation enthalpy defines an allowed range of values for the growth conditions of IrMn\textsubscript{3}; thus, avoiding the synthesis of undesirable phases. We calculated values for $\mu_{Ir}^{bulk} = - 8.865$ eV, $\mu_{Mn}^{bulk} = -9.013$ eV and $\mu_{IrMn\textsubscript{3}}^{bulk} = -37.195$ eV. Thus, using the formation enthalpy equation we find that $\Delta H_f = 1.29$ eV. The range of numerical values for $\Delta\mu_{IrMn\textsubscript{3}}$ is delimited by:

\begin{equation} \label{eq:8}
    -\Delta H_f \leq \Delta\mu_{IrMn\textsubscript{3}} \leq \Delta H_f
\end{equation}

This allows us to plot the formation energies in a range of synthesis conditions that extend from Ir-rich/Mn-poor ($3\mu_{Mn}-\mu_{Ir}-\Delta H_f$) to Ir-poor/Mn-rich ($3\mu_{Mn}-\mu_{Ir}+\Delta H_f$), while avoiding bulk Ir/Mn phases. 

\subsubsection*{Fe surface}
In order to evaluate the formation of the IrMn\textsubscript{3}/Fe interface, the growth of the Fe surface in varying growth conditions must be taken into account. We can achieve this through a similar formalism as the one presented for the IrMn\textsubscript{3} surface. In this case, we only have one type of atomic species; therefore, the surface formation energy can be expressed through the following equation

\begin{equation}\label{eq:9}
    \sigma_{Fe} = \frac{1}{2A}(E_{Fe}^{slab} - n_{Fe}\Delta\mu_{Fe})
\end{equation}

where $A$ is the cross-sectional Fe surface area and the $\frac{1}{2}$ factor comes from taking into account two identical terminations (above and below). $E_{Fe}^{slab}$ is the surface total energy obtained directly from the DFT calculations, $n_{Fe}$ is the number of Fe atoms in the surface and $\Delta\mu_{Fe}$ takes values that range from $\mu_{Fe}$ (Fe-rich conditions) to $\mu_{Fe} - E_{Fe}^{coh}$ (Fe-poor conditions). $E_{Fe}^{coh}$ is the cohesive energy for iron, which corresponds to the difference between the energy of an atom in the bulk structure and the energy of a free atom. 

\subsection{Interface Formation Energy}
\subsubsection*{IrMn\textsubscript{3}/Fe Heterostructure}
The total energy of the IrMn\textsubscript{3}/Fe heterostructure can be expressed as follows

\begin{equation}\label{eq:10}
    E_{IrMn\textsubscript{3}/Fe}^{het} = ( n_{Ir}\mu_{Ir} + n_{Mn}\mu_{Mn} ) + n_{Fe}\mu_{Fe} + A\Gamma + A\sigma_{Fe}
\end{equation}

Where $A$ is the heterostructure cross-sectional area, $Gamma$ is the interface formation energy, and $\sigma_{Fe}$ is the Fe surface formation energy. The terms $n_i$ and $mu_i$ are the number of atoms and chemical potential of each ith  species in the structure. We introduce a term $\Delta E$, which links the total energy of the heterostructure with the energies of the IrMn\textsubscript{3} and Fe surfaces.

\begin{equation}\label{eq:11}
    \Delta E = E_{IrMn\textsubscript{3}/Fe}^{het} - ( E_{IrMn\textsubscript{3}} + E_{Fe} )
\end{equation}

where the total energies of each surface are given by:

\begin{equation}\label{eq:12}
    E_{IrMn\textsubscript{3}} =  n_{Ir}\mu_{Ir} + n_{Mn}\mu_{Mn} + A\sigma_{IrMn\textsubscript{3}}
\end{equation}

    and
    
\begin{equation}\label{eq:13}
    E_{Fe} = n_{Fe}\mu_{Fe} + 2A\sigma_{Fe} .
\end{equation}

The terms $\sigma_{IrMn\textsubscript{3}}$ and $\sigma_{Fe}$ are defined in Eqs. \ref{eq:6}, \ref{eq:7} and \ref{eq:9}. Now, from Eqs. \ref{eq:10}-\ref{eq:13}, we can rewrite the interface formation energy in terms of $\Delta E$:

\begin{equation}\label{eq:14}
    \Gamma = \frac{\Delta E}{A} + \sigma_{IrMn\textsubscript{3}} + \sigma_{Fe}
\end{equation}

where $\Delta\mu_{IrMn\textsubscript{3}}$ allows for variation in the growth conditions of Ir and Mn, and $\Delta \mu_{Fe}$ allows for variation in the growth conditions of Fe. The results are plotted as three dimensional planes, where the $x$ and $y$ axes represent the variation in synthesis conditions, i.e. $\Delta\mu_{IrMn\textsubscript{3}}$ and $\Delta\mu_{Fe}$, while the $z$ axis represents the interface formation energy.


\section{Acknowledgements}

The authors thank DGAPA-UNAM projects IN101019 and
IA100920 and Conacyt grant A1-S-9070 for partial financial support. Calculations were performed in the DGCTIC-UNAM supercomputing center project LANCAD-UNAMDGTIC-051 and LANCAD-UNAM-DGTIC-368. The authors thankfully acknowledge the computer resources, technical expertise and support
provided by the Laboratorio Nacional de Supercómputo del Sureste de México, CONACYT member of the network of national laboratories. J.G.-S. acknowledges THUBAT
KAAL IPICYT supercomputing center for computational
resources. D.M.-L. thanks professor Amit Kohn for providing experimental TEM data. The authors acknowledge E. Murillo and A. Rodriguez-Guerrero for technical assistance and useful discussions.



\section{Supplementary Material}

Table listing the magnetic moments of IrMn-b2 and MnMn-b2 optimized surface structures. Table listing Mean Angular Differences between IrMn\textsubscript{3}/Fe magnetic moments near the surface and pristine T1-AF structure. Figures for: (1) proposed surface terminations, (2) interface formation energy planes, and (3) magnetic coupling behavior in IrMn\textsubscript{3}/Fe heterostructures with Fe magnetization directions Fe[001], Fe[00-1], Fe[110] and Fe[-1-10].

\bibliography{main}

\providecommand{\latin}[1]{#1}
\makeatletter
\providecommand{\doi}
  {\begingroup\let\do\@makeother\dospecials
  \catcode`\{=1 \catcode`\}=2 \doi@aux}
\providecommand{\doi@aux}[1]{\endgroup\texttt{#1}}
\makeatother
\providecommand*\mcitethebibliography{\thebibliography}
\csname @ifundefined\endcsname{endmcitethebibliography}
  {\let\endmcitethebibliography\endthebibliography}{}
\begin{mcitethebibliography}{37}
\providecommand*\natexlab[1]{#1}
\providecommand*\mciteSetBstSublistMode[1]{}
\providecommand*\mciteSetBstMaxWidthForm[2]{}
\providecommand*\mciteBstWouldAddEndPuncttrue
  {\def\EndOfBibitem{\unskip.}}
\providecommand*\mciteBstWouldAddEndPunctfalse
  {\let\EndOfBibitem\relax}
\providecommand*\mciteSetBstMidEndSepPunct[3]{}
\providecommand*\mciteSetBstSublistLabelBeginEnd[3]{}
\providecommand*\EndOfBibitem{}
\mciteSetBstSublistMode{f}
\mciteSetBstMaxWidthForm{subitem}{(\alph{mcitesubitemcount})}
\mciteSetBstSublistLabelBeginEnd
  {\mcitemaxwidthsubitemform\space}
  {\relax}
  {\relax}

\bibitem[Nogu{\'e}s and Schuller(1999)Nogu{\'e}s, and
  Schuller]{nogues1999exchange}
Nogu{\'e}s,~J.; Schuller,~I.~K. Exchange bias. \emph{Journal of Magnetism and
  Magnetic Materials} \textbf{1999}, \emph{192}, 203--232\relax
\mciteBstWouldAddEndPuncttrue
\mciteSetBstMidEndSepPunct{\mcitedefaultmidpunct}
{\mcitedefaultendpunct}{\mcitedefaultseppunct}\relax
\EndOfBibitem
\bibitem[Meiklejohn and Bean(1956)Meiklejohn, and Bean]{meiklejohn1956new}
Meiklejohn,~W.~H.; Bean,~C.~P. New magnetic anisotropy. \emph{Physical review}
  \textbf{1956}, \emph{102}, 1413\relax
\mciteBstWouldAddEndPuncttrue
\mciteSetBstMidEndSepPunct{\mcitedefaultmidpunct}
{\mcitedefaultendpunct}{\mcitedefaultseppunct}\relax
\EndOfBibitem
\bibitem[Stiles and McMichael(2001)Stiles, and McMichael]{stiles2001coercivity}
Stiles,~M.~D.; McMichael,~R.~D. Coercivity in exchange-bias bilayers.
  \emph{Physical review B} \textbf{2001}, \emph{63}, 064405\relax
\mciteBstWouldAddEndPuncttrue
\mciteSetBstMidEndSepPunct{\mcitedefaultmidpunct}
{\mcitedefaultendpunct}{\mcitedefaultseppunct}\relax
\EndOfBibitem
\bibitem[Fernandez-Outon \latin{et~al.}(2008)Fernandez-Outon, O'Grady, Oh,
  Zhou, and Pakala]{fernandez2008large}
Fernandez-Outon,~L.~E.; O'Grady,~K.; Oh,~S.; Zhou,~M.; Pakala,~M. Large
  exchange bias {IrMn/CoFe} for magnetic tunnel junctions. \emph{IEEE
  Transactions on Magnetics} \textbf{2008}, \emph{44}, 2824--2827\relax
\mciteBstWouldAddEndPuncttrue
\mciteSetBstMidEndSepPunct{\mcitedefaultmidpunct}
{\mcitedefaultendpunct}{\mcitedefaultseppunct}\relax
\EndOfBibitem
\bibitem[Pan and Shen(2005)Pan, and Shen]{pan2005magnetically}
Pan,~C.; Shen,~S.-C. Magnetically actuated bi-directional microactuators with
  permalloy and {Fe/Pt} hard magnet. \emph{Journal of magnetism and magnetic
  materials} \textbf{2005}, \emph{285}, 422--432\relax
\mciteBstWouldAddEndPuncttrue
\mciteSetBstMidEndSepPunct{\mcitedefaultmidpunct}
{\mcitedefaultendpunct}{\mcitedefaultseppunct}\relax
\EndOfBibitem
\bibitem[Coehoorn(2000)]{coehoorn2000giant}
Coehoorn,~R. \emph{Magnetic Multilayers and Giant Magnetoresistance}; Springer,
  2000; pp 65--127\relax
\mciteBstWouldAddEndPuncttrue
\mciteSetBstMidEndSepPunct{\mcitedefaultmidpunct}
{\mcitedefaultendpunct}{\mcitedefaultseppunct}\relax
\EndOfBibitem
\bibitem[Lacour \latin{et~al.}(2002)Lacour, Jaffrès, Nguyen Van~Dau, Petroff,
  Vaurès, and Humbert]{doi:10.1063/1.1450050}
Lacour,~D.; Jaffrès,~H.; Nguyen Van~Dau,~F.; Petroff,~F.; Vaurès,~A.;
  Humbert,~J. Field sensing using the magnetoresistance of {IrMn}
  exchange-biased tunnel junctions. \emph{Journal of Applied Physics}
  \textbf{2002}, \emph{91}, 4655--4658\relax
\mciteBstWouldAddEndPuncttrue
\mciteSetBstMidEndSepPunct{\mcitedefaultmidpunct}
{\mcitedefaultendpunct}{\mcitedefaultseppunct}\relax
\EndOfBibitem
\bibitem[Freitas \latin{et~al.}(1994)Freitas, Leal, Melo, Oliveira, Rodrigues,
  and Sousa]{doi:10.1063/1.112304}
Freitas,~P.~P.; Leal,~J.~L.; Melo,~L.~V.; Oliveira,~N.~J.; Rodrigues,~L.;
  Sousa,~A.~T. Spin‐valve sensors exchange‐biased by ultrathin {TbCo}
  films. \emph{Applied Physics Letters} \textbf{1994}, \emph{65},
  493--495\relax
\mciteBstWouldAddEndPuncttrue
\mciteSetBstMidEndSepPunct{\mcitedefaultmidpunct}
{\mcitedefaultendpunct}{\mcitedefaultseppunct}\relax
\EndOfBibitem
\bibitem[Béa \latin{et~al.}(2006)Béa, Bibes, Cherifi, Nolting, Warot-Fonrose,
  Fusil, Herranz, Deranlot, Jacquet, Bouzehouane, and
  Barthélémy]{doi:10.1063/1.2402204}
Béa,~H.; Bibes,~M.; Cherifi,~S.; Nolting,~F.; Warot-Fonrose,~B.; Fusil,~S.;
  Herranz,~G.; Deranlot,~C.; Jacquet,~E.; Bouzehouane,~K.; Barthélémy,~A.
  Tunnel magnetoresistance and robust room temperature exchange bias with
  multiferroic {BiFeO\textsubscript{3}} epitaxial thin films. \emph{Applied
  Physics Letters} \textbf{2006}, \emph{89}, 242114\relax
\mciteBstWouldAddEndPuncttrue
\mciteSetBstMidEndSepPunct{\mcitedefaultmidpunct}
{\mcitedefaultendpunct}{\mcitedefaultseppunct}\relax
\EndOfBibitem
\bibitem[Anderson \latin{et~al.}(2000)Anderson, Huai, and
  Miloslawsky]{doi:10.1063/1.372907}
Anderson,~G.; Huai,~Y.; Miloslawsky,~L. {CoFe/IrMn} exchange biased top,
  bottom, and dual spin valves. \emph{Journal of Applied Physics}
  \textbf{2000}, \emph{87}, 6989--6991\relax
\mciteBstWouldAddEndPuncttrue
\mciteSetBstMidEndSepPunct{\mcitedefaultmidpunct}
{\mcitedefaultendpunct}{\mcitedefaultseppunct}\relax
\EndOfBibitem
\bibitem[{Ikeda} \latin{et~al.}(2007){Ikeda}, {Hayakawa}, {Lee}, {Matsukura},
  {Ohno}, {Hanyu}, and {Ohno}]{4160113}
{Ikeda},~S.; {Hayakawa},~J.; {Lee},~Y.~M.; {Matsukura},~F.; {Ohno},~Y.;
  {Hanyu},~T.; {Ohno},~H. Magnetic Tunnel Junctions for Spintronic Memories and
  Beyond. \emph{IEEE Transactions on Electron Devices} \textbf{2007},
  \emph{54}, 991--1002\relax
\mciteBstWouldAddEndPuncttrue
\mciteSetBstMidEndSepPunct{\mcitedefaultmidpunct}
{\mcitedefaultendpunct}{\mcitedefaultseppunct}\relax
\EndOfBibitem
\bibitem[Sakuma \latin{et~al.}(2003)Sakuma, Fukamichi, Sasao, and
  Umetsu]{sakuma2003first}
Sakuma,~A.; Fukamichi,~K.; Sasao,~K.; Umetsu,~R. First-principles study of the
  magnetic structures of ordered and disordered {Mn-Ir} alloys. \emph{Physical
  Review B} \textbf{2003}, \emph{67}, 024420\relax
\mciteBstWouldAddEndPuncttrue
\mciteSetBstMidEndSepPunct{\mcitedefaultmidpunct}
{\mcitedefaultendpunct}{\mcitedefaultseppunct}\relax
\EndOfBibitem
\bibitem[Tomeno \latin{et~al.}(1999)Tomeno, Fuke, Iwasaki, Sahashi, and
  Tsunoda]{doi:10.1063/1.371298}
Tomeno,~I.; Fuke,~H.~N.; Iwasaki,~H.; Sahashi,~M.; Tsunoda,~Y. Magnetic neutron
  scattering study of ordered {Mn\textsubscript{3}Ir}. \emph{Journal of Applied
  Physics} \textbf{1999}, \emph{86}, 3853--3856\relax
\mciteBstWouldAddEndPuncttrue
\mciteSetBstMidEndSepPunct{\mcitedefaultmidpunct}
{\mcitedefaultendpunct}{\mcitedefaultseppunct}\relax
\EndOfBibitem
\bibitem[Yamaoka(1974)]{doi:10.1143/JPSJ.36.445}
Yamaoka,~T. Antiferromagnetism in $\gamma$-Phase {Mn-Ir} Alloys. \emph{Journal
  of the Physical Society of Japan} \textbf{1974}, \emph{36}, 445--450\relax
\mciteBstWouldAddEndPuncttrue
\mciteSetBstMidEndSepPunct{\mcitedefaultmidpunct}
{\mcitedefaultendpunct}{\mcitedefaultseppunct}\relax
\EndOfBibitem
\bibitem[Kohn \latin{et~al.}(2013)Kohn, Kov{\'a}cs, Fan, McIntyre, Ward, and
  Goff]{kohn2013antiferromagnetic}
Kohn,~A.; Kov{\'a}cs,~A.; Fan,~R.; McIntyre,~G.; Ward,~R.; Goff,~J. The
  antiferromagnetic structures of {IrMn\textsubscript{3}} and their influence
  on exchange-bias. \emph{Scientific reports} \textbf{2013}, \emph{3},
  2412\relax
\mciteBstWouldAddEndPuncttrue
\mciteSetBstMidEndSepPunct{\mcitedefaultmidpunct}
{\mcitedefaultendpunct}{\mcitedefaultseppunct}\relax
\EndOfBibitem
\bibitem[Zhang \latin{et~al.}(2016)Zhang, Han, Yang, Sun, Zhang, Yan, and
  Parkin]{Zhange1600759}
Zhang,~W.; Han,~W.; Yang,~S.-H.; Sun,~Y.; Zhang,~Y.; Yan,~B.; Parkin,~S. S.~P.
  Giant facet-dependent spin-orbit torque and spin Hall conductivity in the
  triangular antiferromagnet {IrMn\textsubscript{3}}. \emph{Science Advances}
  \textbf{2016}, \emph{2}\relax
\mciteBstWouldAddEndPuncttrue
\mciteSetBstMidEndSepPunct{\mcitedefaultmidpunct}
{\mcitedefaultendpunct}{\mcitedefaultseppunct}\relax
\EndOfBibitem
\bibitem[Sharma \latin{et~al.}(2019)Sharma, Hoffmann, Matthes, Busse,
  Selyshchev, Mack, Exner, Horn, Schulz, Zahn, and Salvan]{SHARMA2019165390}
Sharma,~A.; Hoffmann,~M.; Matthes,~P.; Busse,~S.; Selyshchev,~O.; Mack,~P.;
  Exner,~H.; Horn,~A.; Schulz,~S.; Zahn,~D.; Salvan,~G. Exchange bias and
  diffusion processes in laser annealed {CoFeB/IrMn} thin films. \emph{Journal
  of Magnetism and Magnetic Materials} \textbf{2019}, \emph{489}, 165390\relax
\mciteBstWouldAddEndPuncttrue
\mciteSetBstMidEndSepPunct{\mcitedefaultmidpunct}
{\mcitedefaultendpunct}{\mcitedefaultseppunct}\relax
\EndOfBibitem
\bibitem[Jara \latin{et~al.}(2016)Jara, Barsukov, Youngblood, Chen, Read, Chen,
  Braganca, and Krivorotov]{JaraArticle}
Jara,~A.; Barsukov,~I.; Youngblood,~B.; Chen,~Y.-J.; Read,~J.; Chen,~H.;
  Braganca,~P.; Krivorotov,~I. {Highly Textured IrMn\textsubscript{3}(111) Thin
  Films Grown by Magnetron Sputtering}. \emph{IEEE Magnetics Letters}
  \textbf{2016}, \emph{7}, 1--1\relax
\mciteBstWouldAddEndPuncttrue
\mciteSetBstMidEndSepPunct{\mcitedefaultmidpunct}
{\mcitedefaultendpunct}{\mcitedefaultseppunct}\relax
\EndOfBibitem
\bibitem[Jenkins \latin{et~al.}(2019)Jenkins, Chantrell, Klemmer, and
  Evans]{8f1735622d0c42f99d531bb0b0069f94}
Jenkins,~S.; Chantrell,~R.; Klemmer,~T.; Evans,~R. Magnetic anisotropy of the
  noncollinear antiferromagnet {IrMn\textsubscript{3}}. \emph{Physical Review
  B} \textbf{2019}, \emph{100}\relax
\mciteBstWouldAddEndPuncttrue
\mciteSetBstMidEndSepPunct{\mcitedefaultmidpunct}
{\mcitedefaultendpunct}{\mcitedefaultseppunct}\relax
\EndOfBibitem
\bibitem[Szunyogh \latin{et~al.}(2011)Szunyogh, Udvardi, Jackson, Nowak, and
  Chantrell]{articleSzunyogh2011}
Szunyogh,~L.; Udvardi,~L.; Jackson,~J.; Nowak,~U.; Chantrell,~R. Atomistic spin
  model based on a spin-cluster expansion technique: Application to the
  {IrMn${}_{3}$/Co} interface. \emph{Phys. Rev. B} \textbf{2011}, \emph{83},
  024401\relax
\mciteBstWouldAddEndPuncttrue
\mciteSetBstMidEndSepPunct{\mcitedefaultmidpunct}
{\mcitedefaultendpunct}{\mcitedefaultseppunct}\relax
\EndOfBibitem
\bibitem[Yanes \latin{et~al.}(2013)Yanes, Jackson, Udvardi, Szunyogh, and
  Nowak]{yanes2013exchange}
Yanes,~R.; Jackson,~J.; Udvardi,~L.; Szunyogh,~L.; Nowak,~U. Exchange bias
  driven by {Dzyaloshinskii-Moriya} interactions. \emph{Physical review
  letters} \textbf{2013}, \emph{111}, 217202\relax
\mciteBstWouldAddEndPuncttrue
\mciteSetBstMidEndSepPunct{\mcitedefaultmidpunct}
{\mcitedefaultendpunct}{\mcitedefaultseppunct}\relax
\EndOfBibitem
\bibitem[Guerrero-S{\'a}nchez \latin{et~al.}(2015)Guerrero-S{\'a}nchez, Mandru,
  Wang, Takeuchi, Cocoletzi, and Smith]{guerrero2015structural}
Guerrero-S{\'a}nchez,~J.; Mandru,~A.-O.; Wang,~K.; Takeuchi,~N.;
  Cocoletzi,~G.~H.; Smith,~A.~R. Structural, electronic and magnetic properties
  of {Mn3N2(0 0 1)} surfaces. \emph{Applied Surface Science} \textbf{2015},
  \emph{355}, 623--630\relax
\mciteBstWouldAddEndPuncttrue
\mciteSetBstMidEndSepPunct{\mcitedefaultmidpunct}
{\mcitedefaultendpunct}{\mcitedefaultseppunct}\relax
\EndOfBibitem
\bibitem[Liu \latin{et~al.}(2005)Liu, Chizmeshya, Kouvetakis, and
  Tsong]{liu2005first}
Liu,~P.-L.; Chizmeshya,~A.; Kouvetakis,~J.; Tsong,~I. First-principles studies
  of {GaN}(0001) heteroepitaxy on {ZrB\textsubscript{2}}(0001). \emph{Physical
  Review B} \textbf{2005}, \emph{72}, 245335\relax
\mciteBstWouldAddEndPuncttrue
\mciteSetBstMidEndSepPunct{\mcitedefaultmidpunct}
{\mcitedefaultendpunct}{\mcitedefaultseppunct}\relax
\EndOfBibitem
\bibitem[Dong \latin{et~al.}(2011)Dong, Zhang, Yunoki, Liu, and
  Dagotto]{dong2011ab}
Dong,~S.; Zhang,~Q.; Yunoki,~S.; Liu,~J.-M.; Dagotto,~E. Ab initio study of the
  intrinsic exchange bias at the
  {SrRuO\textsubscript{3}/SrMnO\textsubscript{3}} interface. \emph{Physical
  Review B} \textbf{2011}, \emph{84}, 224437\relax
\mciteBstWouldAddEndPuncttrue
\mciteSetBstMidEndSepPunct{\mcitedefaultmidpunct}
{\mcitedefaultendpunct}{\mcitedefaultseppunct}\relax
\EndOfBibitem
\bibitem[Jadaun \latin{et~al.}(2020)Jadaun, Register, and
  Banerjee]{jadaun2020microscopic}
Jadaun,~P.; Register,~L.~F.; Banerjee,~S.~K. The microscopic origin of {DMI} in
  magnetic bilayers and prediction of giant {DMI} in new bilayers. \emph{npj
  Computational Materials} \textbf{2020}, \emph{6}, 1--6\relax
\mciteBstWouldAddEndPuncttrue
\mciteSetBstMidEndSepPunct{\mcitedefaultmidpunct}
{\mcitedefaultendpunct}{\mcitedefaultseppunct}\relax
\EndOfBibitem
\bibitem[Gayles \latin{et~al.}(2015)Gayles, Freimuth, Schena, Lani,
  Mavropoulos, Duine, Bl{\"u}gel, Sinova, and
  Mokrousov]{gayles2015dzyaloshinskii}
Gayles,~J.; Freimuth,~F.; Schena,~T.; Lani,~G.; Mavropoulos,~P.; Duine,~R.;
  Bl{\"u}gel,~S.; Sinova,~J.; Mokrousov,~Y. {Dzyaloshinskii-Moriya interaction
  and Hall effects in the skyrmion phase of
  Mn\textsubscript{1-x}Fe\textsubscript{x}Ge}. \emph{Physical review letters}
  \textbf{2015}, \emph{115}, 036602\relax
\mciteBstWouldAddEndPuncttrue
\mciteSetBstMidEndSepPunct{\mcitedefaultmidpunct}
{\mcitedefaultendpunct}{\mcitedefaultseppunct}\relax
\EndOfBibitem
\bibitem[Sort \latin{et~al.}(2005)Sort, Baltz, Garcia, Rodmacq, and
  Dieny]{sort2005tailoring}
Sort,~J.; Baltz,~V.; Garcia,~F.; Rodmacq,~B.; Dieny,~B. Tailoring perpendicular
  exchange bias in {[Pt/Co]-IrMn multilayers}. \emph{Physical Review B}
  \textbf{2005}, \emph{71}, 054411\relax
\mciteBstWouldAddEndPuncttrue
\mciteSetBstMidEndSepPunct{\mcitedefaultmidpunct}
{\mcitedefaultendpunct}{\mcitedefaultseppunct}\relax
\EndOfBibitem
\bibitem[Baltz \latin{et~al.}(2005)Baltz, Sort, Landis, Rodmacq, and
  Dieny]{baltz2005tailoring}
Baltz,~V.; Sort,~J.; Landis,~S.; Rodmacq,~B.; Dieny,~B. Tailoring size effects
  on the exchange bias in ferromagnetic-antiferromagnetic < 100 nm
  nanostructures. \emph{Physical review letters} \textbf{2005}, \emph{94},
  117201\relax
\mciteBstWouldAddEndPuncttrue
\mciteSetBstMidEndSepPunct{\mcitedefaultmidpunct}
{\mcitedefaultendpunct}{\mcitedefaultseppunct}\relax
\EndOfBibitem
\bibitem[Zhou \latin{et~al.}(2017)Zhou, Song, Bai, Quan, Jiang, Liu, Xu, Dhesi,
  and Xu]{zhou2017robust}
Zhou,~G.; Song,~C.; Bai,~Y.; Quan,~Z.; Jiang,~F.; Liu,~W.; Xu,~Y.;
  Dhesi,~S.~S.; Xu,~X. Robust interfacial exchange bias and metal--insulator
  transition influenced by the {LaNiO\textsubscript{3}} layer thickness in
  {La\textsubscript{0.7}Sr\textsubscript{0.3}MnO\textsubscript{3}/LaNiO\textsubscript{3}}
  superlattices. \emph{ACS applied materials \& interfaces} \textbf{2017},
  \emph{9}, 3156--3160\relax
\mciteBstWouldAddEndPuncttrue
\mciteSetBstMidEndSepPunct{\mcitedefaultmidpunct}
{\mcitedefaultendpunct}{\mcitedefaultseppunct}\relax
\EndOfBibitem
\bibitem[Liu \latin{et~al.}(2001)Liu, Baker, Tuominen, Russell, and
  Schuller]{liu2001tailoring}
Liu,~K.; Baker,~S.~M.; Tuominen,~M.; Russell,~T.~P.; Schuller,~I.~K. Tailoring
  exchange bias with magnetic nanostructures. \emph{Physical Review B}
  \textbf{2001}, \emph{63}, 060403\relax
\mciteBstWouldAddEndPuncttrue
\mciteSetBstMidEndSepPunct{\mcitedefaultmidpunct}
{\mcitedefaultendpunct}{\mcitedefaultseppunct}\relax
\EndOfBibitem
\bibitem[Guerrero-S{\'a}nchez and Takeuchi(2018)Guerrero-S{\'a}nchez, and
  Takeuchi]{guerrero2018formation}
Guerrero-S{\'a}nchez,~J.; Takeuchi,~N. Formation of ferromagnetic/ferrimagnetic
  epitaxial interfaces: Stability and magnetic properties. \emph{Computational
  Materials Science} \textbf{2018}, \emph{144}, 294--303\relax
\mciteBstWouldAddEndPuncttrue
\mciteSetBstMidEndSepPunct{\mcitedefaultmidpunct}
{\mcitedefaultendpunct}{\mcitedefaultseppunct}\relax
\EndOfBibitem
\bibitem[Corbett \latin{et~al.}(2019)Corbett, Guerrero-Sanchez, Gallagher,
  Mandru, Richard, Ingram, Yang, Takeuchi, and Smith]{corbett2019dislocation}
Corbett,~J.~P.; Guerrero-Sanchez,~J.; Gallagher,~J.~C.; Mandru,~A.-O.;
  Richard,~A.~L.; Ingram,~D.~C.; Yang,~F.; Takeuchi,~N.; Smith,~A.~R.
  Dislocation structures, interfacing, and magnetism in the
  {L1\textsubscript{0}-MnGa on $\eta
  \bot$-Mn\textsubscript{3}N\textsubscript{2}} bilayer. \emph{Journal of Vacuum
  Science \& Technology A} \textbf{2019}, \emph{37}, 031102\relax
\mciteBstWouldAddEndPuncttrue
\mciteSetBstMidEndSepPunct{\mcitedefaultmidpunct}
{\mcitedefaultendpunct}{\mcitedefaultseppunct}\relax
\EndOfBibitem
\bibitem[Kresse and Furthmüller(1996)Kresse, and
  Furthmüller]{PhysRevB.54.11169}
Kresse,~G.; Furthmüller,~J. Efficient iterative schemes for ab initio
  total-energy calculations using a plane-wave basis set. \emph{Phys. Rev. B}
  \textbf{1996}, \emph{54}, 11169--11186\relax
\mciteBstWouldAddEndPuncttrue
\mciteSetBstMidEndSepPunct{\mcitedefaultmidpunct}
{\mcitedefaultendpunct}{\mcitedefaultseppunct}\relax
\EndOfBibitem
\bibitem[Bl{\"o}chl(1994)]{blochl1994projector}
Bl{\"o}chl,~P.~E. Projector augmented-wave method. \emph{Physical review B}
  \textbf{1994}, \emph{50}, 17953\relax
\mciteBstWouldAddEndPuncttrue
\mciteSetBstMidEndSepPunct{\mcitedefaultmidpunct}
{\mcitedefaultendpunct}{\mcitedefaultseppunct}\relax
\EndOfBibitem
\bibitem[Kresse and Joubert(1999)Kresse, and Joubert]{kresse1999ultrasoft}
Kresse,~G.; Joubert,~D. From ultrasoft pseudopotentials to the projector
  augmented-wave method. \emph{Physical review B} \textbf{1999}, \emph{59},
  1758\relax
\mciteBstWouldAddEndPuncttrue
\mciteSetBstMidEndSepPunct{\mcitedefaultmidpunct}
{\mcitedefaultendpunct}{\mcitedefaultseppunct}\relax
\EndOfBibitem
\bibitem[Perdew \latin{et~al.}(1996)Perdew, Burke, and
  Ernzerhof]{perdew1996generalized}
Perdew,~J.~P.; Burke,~K.; Ernzerhof,~M. Generalized gradient approximation made
  simple. \emph{Physical review letters} \textbf{1996}, \emph{77}, 3865\relax
\mciteBstWouldAddEndPuncttrue
\mciteSetBstMidEndSepPunct{\mcitedefaultmidpunct}
{\mcitedefaultendpunct}{\mcitedefaultseppunct}\relax
\EndOfBibitem
\end{mcitethebibliography}

\newpage
\section{TOC Figure}
\begin{figure}
    \centering
    \includegraphics[width=12cm]{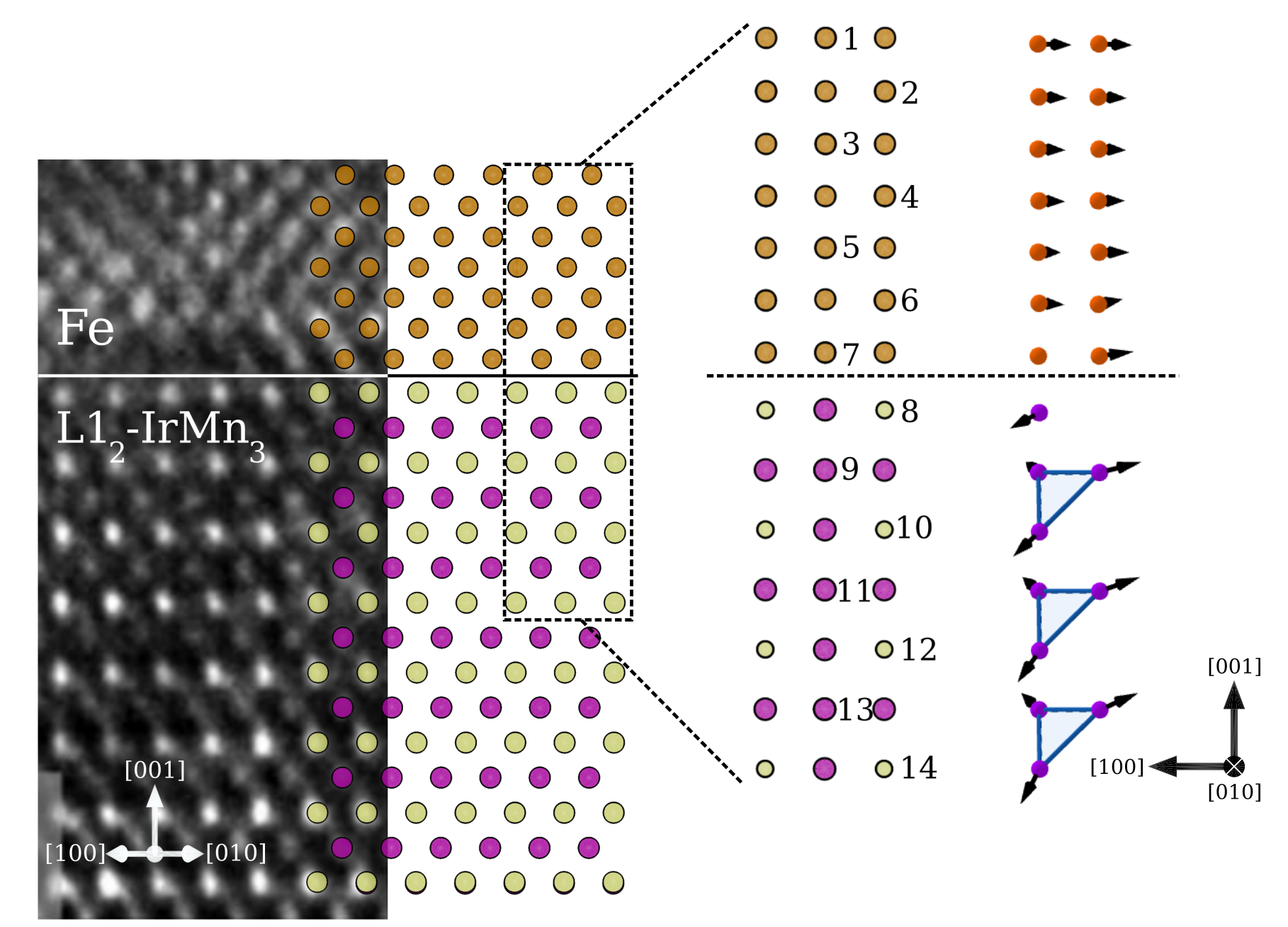}
\end{figure}

\end{document}